\documentclass[11pt,onecolumn]{IEEEtran}

\usepackage{amsmath}
\usepackage{amssymb}
\usepackage{graphicx}
\usepackage{color}
\usepackage{subfig}

\newtheorem{lemma}{Lemma}
\newtheorem{theorem}{Theorem}

\newcommand{\GoesTo}{\longrightarrow}
\newcommand{\Equiv}{\Leftrightarrow}
\newcommand{\To}{\longrightarrow}

\begin{document}
\title{Selfish Response to Epidemic Propagation \thanks{This work has been submitted to the IEEE for possible
publication. Copyright may be transferred without notice, after
which this version may no longer be accessible.}}

\author{George~Theodorakopoulos,
        Jean-Yves~Le~Boudec,~\IEEEmembership{Fellow,~IEEE,}
        and~John~S.~Baras,~\IEEEmembership{Fellow,~IEEE}
        \thanks{George Theodorakopoulos (corresponding author -- george.theodorakopoulos @ epfl.ch) and Jean-Yves Le Boudec (jean-yves.leboudec @ epfl.ch) are with the Ecole Polytechnique Federale de Lausanne, EPFL-IC-LCA, Building BC, Station 14, CH-1015 Lausanne, Switzerland. John S. Baras (baras @ umd.edu) is with the University of Maryland, Department of Electrical \& Computer Engineering, 2247 AV Williams Building, College Park, MD 20742-3271.}}

\maketitle

\begin{abstract}
  An epidemic spreading in a network calls for a decision on the part of the network members: They should decide whether to protect themselves or not. Their decision depends on the trade-off between their perceived risk of being infected and the cost of being protected. The network members can make decisions repeatedly, based on information that they receive about the changing infection level in the network.

  We study the equilibrium states reached by a network whose members increase (resp. decrease) their security deployment when learning that the network infection is widespread (resp. limited). Our main finding is that the equilibrium level of infection increases as the learning rate of the members increases. We confirm this result in three scenarios for the behavior of the members: strictly rational cost minimizers, not strictly rational, and strictly rational but split into two response classes. In the first two cases, we completely characterize the stability and the domains of attraction of the equilibrium points, even though the first case leads to a differential inclusion. We validate our conclusions with simulations on human mobility traces.
\end{abstract}

\section{Introduction}
Epidemiology research has made extensive use of disease spreading models to study how a virus propagates in a human population \cite{hethcote-infectious}. Shortly after the appearance of self-replicating malicious programs in computers, aptly named \emph{computer viruses}, security researchers turned to epidemic models to study the propagation of these programs \cite{kephart-directed, kephart-measuring}. More recently, the proliferation of capable mobile devices, such as smartphones, made mobile networks a fertile ground for spreading malware \cite{hypponen-sciam, mobileviruslist}. The propagation characteristics of malware in such networks have been studied and countermeasures have been proposed \cite{terzis-accesspoint,fleizach-addressbook,carreras-eigenvector,zheng-epidemic,khouzani-quarantining,alpcan-epidemic}.

Countermeasures to an infection can be centrally enforced, or the decision for their adoption can be left to individual agents such as individual home computer users, companies, or people in a society. Centralized enforcing is more likely to work in tightly controlled environments, such as within a company network where the users are obliged to abide by the company security policy. However, when it is up to individual agents to invest in protection against infection \cite{jiang-efficiency,lelarge-externalities,aspnes-inoculation,moscibroda-selfish}, there appear contradicting incentives. Although agents want to be safe against real or virtual viruses, they would prefer to avoid investing in security: Security not only costs money, but it usually also reduces the utility of the network by, for example, isolating the agent from the rest of the network, or it reduces the utility of the device by, for example, slowing it down \cite{yan-slowdown}. Another counter-incentive is that the security of a network agent exhibits \emph{positive externalities} with respect to the decisions of others: If others patch their computers, everyone becomes more secure, even those who do not patch their own computer. If others are vaccinated, everyone becomes safer, even those who are not vaccinated. Therefore, agents have an incentive to free-ride on the security investments of others, reaping the benefits without paying the costs. More background on computer network security and individual incentives can be found in two recent books \cite{alpcan-book, buttyan-book}.

However, to the best of our knowledge, only static incentives of agents have been studied: Users are modeled as only making a once-and-for-all decision to install or not a security product. Herein lies our goal:
Agents do not choose only once whether they will invest in security or not. They balance between the cost and benefit of their investment, and the cost and benefit typically change with time. Security advisories exist about current and newly emerging threats in popular technology products \cite{advisory-cert, mobileviruslist}, and about current and newly emerging human epidemics \cite{advisory-who}. We study how changing incentives influence the security decisions and the resulting infection level in the network.

We model agents as more likely to invest in security when the infection level is high and as less likely to invest, or more likely to divest, when the infection level is low. On the one hand, if people receive news of an ongoing epidemic, they are much more willing to protect themselves: the prevalence of AIDS has been observed to increase risk avoidance, either through increased condom use \cite{ahituv-responsiveness}, or through behavioral risk avoidance \cite{stoneburner-behavioral}.

On the other hand, when the infection has subsided and there is no clear danger, complacency may set in with a consequent reduction in the efforts and capital expended to ensure safety. Neglect of human epidemics that are at the point of near extinction has led to their resurgence \cite{gubler-resurgent}. Also, companies and users are not easily convinced to buy security products \cite{schneier-sell}.
If the threat is not extremely clear or imminent (``One solution [for selling security] is to stoke fear.'' \cite{schneier-sell}), users will resist investing in security, they will divest, or they will stop renewing their investment.

In human epidemic modeling it has already been recommended \cite{ferguson-human} to incorporate into models the changing behavior of humans towards protection against ongoing epidemics. In a recently proposed model \cite{funk-awareness}, the awareness for the epidemic spreads in parallel with the disease itself. In particular, the awareness is spread from the aware to the unaware part of the population at some rate and then lost again or forgotten at a different rate. Aware users are less likely to contract the disease because, for example, they choose to stay at home. Nevertheless, we note that a user alternates between states of awareness and unawareness mechanically, \emph{without} making the decision himself, so we cannot speak of individual incentives in this case.

In this paper, we model individuals' \emph{changing} responses that depend myopically on the fluctuating infection level in an ongoing epidemic. We combine the epidemic propagation with a game theoretic description of the user behavior into an Ordinary Differential Equation (ODE) model.

We find that the network reaches an endemic equilibrium, that is, an equilibrium where the infection persists. We reach the counterintuitive conclusion that the higher the learning rate (the rate at which users learn what the infection level is), the higher the infection level at the equilibrium. The effect of the learning rate is less pronounced when the users are more conservative, i.e., when they are willing to invest in protection at lower infection levels. These conclusions hold across the various user behavior functions that we model.

When users are strictly rational cost minimizers, leading to discontinuous best response dynamics, our model turns into a system of differential inclusions, which can also be viewed as a switched nonlinear system \cite{liberzon-basic}. We prove that there can be no periodic solutions, there can only be equilibrium points. We characterize the domains of attraction for these points, as well as prove (local) asymptotic stability results. These findings, presented in Section~\ref{sec:discontinuous}, might also be of theoretical interest for switched nonlinear systems, as the bulk of the research on switched systems is focused on the linear case \cite{antsaklis-survey}.

To account for users who are not strictly rational, we study behavior functions that are continuous. These functions are arbitrary except for the requirement that users be more willing to become and stay protected as the current level of infection increases. We prove system properties that are similar to those in the strictly rational case.

We use simulations on human mobility traces to confirm our main theoretical conclusion that a higher learning rate leads to a higher infection level.

To account for heterogeneity among users, we also study a system with two classes of users (easily extensible to more than two), each with a different sensitivity to the infection level: Users in the first class (Responsible) become protected at lower levels of an infection, whereas users in the second class (Selfish) become protected only at higher levels, thus in a sense free-riding on the security investment of the first class. In this case, too, it holds that a higher learning rate leads to a higher infection level, which we also confirm with simulations.

The remainder of the paper is organized as follows. In Section~\ref{sec:model} we describe our model for the evolution of the network state, comprising an epidemic propagation component and a user behavior component. We study users with a strictly rational behavior (Section~\ref{sec:discontinuous}), then users with non-strictly rational behavior (Section~\ref{sec:continuous}), followed by users with heterogeneous behavior (Section~\ref{sec:multiple}). In Section~\ref{sec:validation} we present an empirical validation of our conclusions through simulations on human mobility traces.

\section{Model for Epidemic Propagation and User Behavior}\label{sec:model}
\subsection{Epidemic Propagation}
  There are $N$ users in the network. Each user can be in one of three states:
  \begin{itemize}
    \item Susceptible, denoted by $S$: The user does not currently deploy security and is not infected.
    \item Infected, denoted by $I$: The user is infected by the virus and will spread it to any susceptible user he makes contact with.
    \item Protected, denoted by $P$: The user deploys security and is therefore immune to the virus.
  \end{itemize}

  The number and fraction of users in each state are denoted, respectively, by $N_S, N_I, N_P$ and $S, I, P$. It follows that $N_S+N_I+N_P=N$ and $S+I+P=1$. The state of the network is $x = (S, I, P)$, and the set of possible states is $X = \frac{1}{N}\mathbb{N}^3 = \{\frac{N_S}{N}, \frac{N_I}{N}, \frac{N_P}{N}\}$.

  The evolution of the network state $x$ is described as a Continuous Time Markov Process, as follows. With each user a Poisson alarm clock of rate $\beta + \gamma +\delta$ is associated. When the clock of user $i$ rings -- say at time $t$ -- one of three events happens:
  \begin{itemize}
    \item[M] With probability $\frac{\beta}{\beta + \gamma +\delta}$, user $i$ has a \emph{meeting} with another user, chosen uniformly at random. If the meeting is between a Susceptible and an Infected user, the Susceptible user becomes Infected. Otherwise nothing happens.
    \item[U] With probability $\frac{\gamma}{\beta + \gamma +\delta}$, user $i$ receives an \emph{update} about the network state $x$, and he has the opportunity to revise his current strategy if his state is $S$ or $P$. If $i$'s state is $S$, he switches to $P$ with probability $p_{SP}(x)$. If $i$'s state is $P$, he switches to $S$ with probability $p_{PS}(x)$. If $i$ is Infected, nothing happens.
    \item[D] With probability $\frac{\delta}{\beta + \gamma +\delta}$, user $i$ has a \emph{disinfection} opportunity. That is, if $i$ is Infected, he becomes disinfected, and we assume he becomes Protected. If $i$ is not Infected, nothing happens.
  \end{itemize}

    Table~\ref{tbl:events} summarizes the possible events and their effect on the network state.
    \begin{table}
    \centering
        \begin{tabular}{|l|l|}
          \hline
          Event & Effect $\Delta x$\\
          \hline
          Meeting between S and I   & $\frac{1}{N}(-1, +1, 0)$\\
          Update of S               & $\frac{1}{N}(-p_{SP}(x), 0, +p_{SP}(x))$\\
          Update of P               & $\frac{1}{N}(+p_{PS}(x), 0, -p_{PS}(x))$\\
          Disinfection of I         & $\frac{1}{N}(0, -1, +1)$\\
          \hline
        \end{tabular}
        \caption{Possible events and their effect on the network state}\label{tbl:events}
    \end{table}

    We consider the large population scenario, i.e., the limit $N\GoesTo\infty$. Kurtz \cite{kurtz-approximation} and Ljung \cite{ljung-stochastic} show that, when $N\GoesTo\infty$, the Continuous Time Markov Process described previously converges to a deterministic function, which is the solution to a system of Ordinary Differential Equations:
    \begin{subequations}\label{eq:mastersystem}
    \begin{align}
      \frac{d}{dt}S& = -\beta SI -\gamma Sp_{SP}(x) + \gamma Pp_{PS}(x)\\
      \frac{d}{dt}I& = \beta SI - \delta I\\
      \frac{d}{dt}P& = \delta I + \gamma Sp_{SP}(x) - \gamma Pp_{PS}(x)
    \end{align}
    \end{subequations}

    Since $S+I+P=1$, we can eliminate one of the three state variables. We eliminate $P$, and the system becomes
    \begin{subequations}\label{eq:mastersystem2d}
    \begin{align}
      \frac{d}{dt}S& = -\beta SI -\gamma Sp_{SP}(x) + \gamma (1-S-I)p_{PS}(x)\\
      \frac{d}{dt}I& = \beta SI - \delta I,
    \end{align}
    \end{subequations}
    together with $P=1-S-I$. The state space is $D=(S,I), 0\leq S, I\leq 1, S+I\leq 1$, and it is bounded. This system is two dimensional and autonomous. Note that for $\gamma=0$, the model is identical to the standard SIR epidemic model \cite{hethcote-infectious} (R stands for Recovered).

   \emph{Remark:} The results of Kurtz and Ljung hold when the resulting deterministic equations are continuous. As we will see when discussing the behavior of users (Sec.~\ref{sec:behavior}), the functions $p_{PS}(x)$ and $p_{SP}(x)$ can also be discontinuous, and indeed multivalued at the discontinuity. Gast and Gaujal \cite{gast-nonsmooth} prove a similar convergence result for the multivalued case: the trajectory of the stochastic system converges in probability to a solution of a differential inclusion, as opposed to a differential equation. If the solution is unique, the stochastic system converges to it so the situation is as in the continuous case. If there are multiple solutions, then the stochastic system can converge to any of them. In the section on the discontinuous dynamics, we will resolve the issue of uniqueness of solutions.

   We will denote the right-hand side of the system~\eqref{eq:mastersystem2d} by $F(x)$, and we will slightly abuse the notation for $x$ to be $x=(S,I), x\in D$. So, the system~\eqref{eq:mastersystem2d} will be written
    \begin{equation}\label{eq:mastersystemabbrv}
      \frac{d}{dt}x = F(x)
    \end{equation}
    for the differential equation, or
    \begin{equation}\label{eq:masterinclusionabbrv}
      \frac{d}{dt}x \in F(x)
    \end{equation}
    for the differential inclusion; the one we refer to will be clear from the context.

    \subsection{User Behavior}\label{sec:behavior}
    As can be seen from the epidemic propagation model, the only point at which the users can make a choice is at an update event. We assume that there is a cost $c_I$ associated with becoming Infected, and a cost $c_P$ associated with becoming Protected. It holds that $c_I > c_P > 0$. There is no cost for being Susceptible. Note that these costs need not be the actual costs; what influences the decisions of users are the costs as perceived by the users.

    If we assume that each user behaves strictly rationally, the choice between Susceptible and Protected depends on which state minimizes the user's expected cost. Specifically, given the aforementioned model of random pair meetings, a user's expected cost at a particular network state $x=(S,I)$ is $c_P$ if he chooses to be Protected and $Ic_I$ if he chooses to be Susceptible, thus risking infection. Therefore, the user's decision would be $S$ if $Ic_I < c_P$, and $P$ if $Ic_I > c_P$. In this case, the functions $p_{SP}(x)$ and $p_{PS}(x)$ would be step functions of $I$:
    \begin{align}\label{eq:br}
      p_{SP}(x) = p_{SP}(I)& = 1\{Ic_I > c_P\}\\
      p_{PS}(x) = p_{PS}(I)& = 1\{Ic_I < c_P\}.
    \end{align}
    If $Ic_I = c_P$, then both choices are optimal, and any randomization between them is also optimal. So, when $Ic_I = c_P$, the functions $p_{SP}(I)$ and $p_{PS}(I)$ are multivalued. For convenience, we define
    \begin{equation}
      I^* \equiv \frac{c_P}{c_I}.
    \end{equation}
    Note that if we were to set $I^*$ to a value larger than 1, then $p_{SP}$ would always be equal to 0, $p_{PS}$ would always be equal to 1, and our model would be identical to the SIRS model \cite{hethcote-infectious}. We revisit this connection when discussing equilibrium points whose $I$-coordinate is less than $I^*$ (Section~\ref{subsec:discont-points}).

    To account for users that cannot be assumed to be strictly rational, or their perception of the cost is not crisp (e.g., they are not sure about the exact values of $c_I$ and $c_P$), or they take the network state report to not be completely accurate, we consider a different scenario for the functions $p_{SP}(\cdot)$ and $p_{PS}(\cdot)$. We assume that they can be arbitrary functions of $I$, as long as the former is non-decreasing with $I$ and the latter is non-increasing with $I$.

    Finally, to account for users with different characteristics, we will consider multiple user classes, each with a different (discontinuous) response function.

    In what follows, first we will consider the case that $p_{SP}(\cdot)$ and $p_{PS}(\cdot)$ are discontinuous step functions and actually multivalued at the discontinuity. Then, we will consider the case that they are continuously differentiable. Last, we will go into the multiple user class scenario.
    
\section{The Users are Strictly Rational}\label{sec:discontinuous}
    The best response correspondence dictates the shape of $p_{SP}(I)$ and $p_{PS}(I)$:
    \begin{equation}
      p_{SP}(I) =\begin{cases}
      0, & I < I^* \\
      [0,1], & I = I^* \\
      1, & I > I^* \\
      \end{cases}
\qquad
      p_{PS}(I) =\begin{cases}
      1, & I < I^* \\
      [0,1], & I = I^* \\
      0, & I > I^* \\
      \end{cases}.
    \end{equation}

    We now have to solve the differential inclusion (recall~\eqref{eq:mastersystem2d} and \eqref{eq:masterinclusionabbrv})
    \begin{equation}\label{eq:masterinclusion}
      \frac{d}{dt}x \in F(x), x\in D.
    \end{equation}
    The vector field of $F(x)$ is plotted in Figure~\ref{fig:discont-vectorfield} for various values of the parameters.

    We define a partition of the state space $D$ into three domains: $D^- = D\cap\{(S, I), I<I^*\}$, $D^+ = D\cap\{(S, I), I>I^*\}$, and $L = D\cap\{(S, I): I=I^*\}$. The domain $L$ will also be referred to as the discontinuity line.

\subsection{Existence of Solutions}
    A solution for this differential inclusion \cite{filippov-book} is an absolutely continuous vector function $x(t)$ defined on an interval $J$ for which $\frac{d}{dt}x(t)\in F(x(t))$ almost everywhere on $J$. From the theory of differential inclusions we know that a solution of \eqref{eq:masterinclusion} exists if, for every $x\in D$, the \emph{basic conditions} apply: The set $F(x)$ is nonempty, bounded, closed, convex, and the function $F$ is upper semi-continuous.

    A set-valued function $f(x)$ is called \emph{upper semi-continuous} at the point $x$ if $\rho(f(x'),f(x))\GoesTo 0$ as $x'\GoesTo x$. The function $\rho(A,B)$ is one characterization of the distance between two nonempty closed sets $A$ and $B$:
    \begin{equation}
      \rho(A,B) = \sup_{a\in A}\inf_{b\in B}d(a,b),
    \end{equation}
    where $d(\cdot,\cdot)$ is the Euclidean distance between two points.

    The basic conditions apply in our case:

    For every $x\notin L$, the set $F(x)$ is a singleton, hence, it is nonempty, bounded, closed, and convex; additionally, the function $F$ is continuous at $x$, hence, it is also upper semi-continuous.

    At each point $x \in L$, the set $F(x)$ is the segment
    \begin{equation}\label{eq:FonL}
      F(S, I^*) = \begin{pmatrix}
                            -\beta SI^* + \gamma[-S, 1-S-I^*]\\
                            \beta SI^* - \delta I^*
                        \end{pmatrix},
    \end{equation}
    which is the smallest convex closed set containing all the limit values of $F(x')$ for $x' \GoesTo x\in L$. When $x' \GoesTo x$ from $D^+$, the limit value of $F(x)$ is
        \begin{equation}
                \begin{pmatrix}
                -\beta SI^* - \gamma S\\
                \beta SI^* - \delta I^*
                \end{pmatrix},
    \end{equation}
    and when $x' \GoesTo x$ from $D^-$, the limit value of $F(x)$ is
        \begin{equation}
                \begin{pmatrix}
                -\beta SI^* + \gamma(1-S-I^*)\\
                \beta SI^* - \delta I^*
                \end{pmatrix}.
    \end{equation}
    The set $F(S, I^*)$ is bounded and upper semi-continuous \cite[Lemma 3, \S 6]{filippov-book}.

\subsection{Uniqueness of Solutions}
    In general, because the right-hand side of \eqref{eq:masterinclusion} is multivalued, even though two solutions at time $t_0$ are both at the point $x_0$, they may not coincide on an interval $t_0\leq t \leq t_1$ for any $t_1>t_0$. If any two solutions that coincide at $t_0$ also coincide until some $t_1>t_0$, then we say that \emph{right uniqueness holds at $(t_0,x_0)$}. Left uniqueness at $(t_0,x_0)$ is defined similarly (with $t_1<t_0$), and (right and left) uniqueness in a domain holds, if it holds at each point of the domain.

    The solution is unique in $D^-$ and in $D^+$ because $F$ has continuous partial derivatives there.

    We will now show when a solution of \eqref{eq:masterinclusion} lying on the line of discontinuity $L$ can be uniquely continued in the direction of increasing $t$. We will see that all solutions can be uniquely continued, except those that start at the point $(S,I)=\left(\frac{\delta}{\beta},I^*\right)$. Those latter solutions all start at the same point and then diverge, but none of them can ever approach that point again in the positive direction of time (the proof is in Lemma~\ref{lem:lemma1}). So, if we ignore the initial point of those solutions, all solutions can be uniquely continued.

    Formally, let $F^-(x)$ and $F^+(x)$ be the limiting values of the function $F$ at a point $x\in L$ as $F$ approaches $x$ from $D^-$ and from $D^+$, respectively. Let $h(x) = F^+(x) - F^-(x)$, and $F_N^-$, $F_N^+$, $h_N$ be the projections of the vectors $F^-$, $F^+$, $h$ onto the vector $n=(0, 1)^T$, the normal to $L$ directed from $D^-$ to $D^+$ at the point $x$.

    The values of these vectors and projections are:
    \begin{align}
      F^-(x)& =(-\beta SI^* + \gamma(1-S-I^*), \beta SI^* - \delta I^*)^T \\
      F^+(x)& =(-\beta SI^* -\gamma S, \beta SI^* - \delta I^*)^T \\
      h(x)& =(-\gamma S - \gamma(1-S-I^*), 0)^T \\
      F_N^-& =\beta SI^* - \delta I^* \\
      F_N^+& =\beta SI^* - \delta I^* \\
      h_N& =0
    \end{align}

    We know \cite[\S10, Corollary 1]{filippov-book} that on the discontinuity line $L$, at the points where $F_N^- > 0$, $F_N^+ > 0$ (or $F_N^- < 0$, $F_N^+ < 0$) the solutions pass from $D^-$ into $D^+$ (correspondingly, from $D^+$ into $D^-$) and uniqueness is not violated. So, at no point of $L$ is uniqueness violated, except at $(S,I)=(\frac{\delta}{\beta},I^*)$.

    A solution that passes from the point $(S,I)=(\frac{\delta}{\beta},I^*)$ will either stay there (if $0 \in F(\frac{\delta}{\beta},I^*)$, i.e., if $(\frac{\delta}{\beta},I^*)$ is an equilibrium point) or it can be continued in multiple ways, all tangent to $L$ as $\frac{d}{dt}I=0$ when $S=\frac{\delta}{\beta}$. Each of these multiple solutions corresponds to a different value of $F(\frac{\delta}{\beta},I^*)$. More details on these trajectories can be found in the proof of Lemma~\ref{lem:lemma1}.

\begin{figure}[ht]
\centering
\subfloat[The case $\delta \geq \beta$. The only equilibrium point is $X_0=(1,0)$. It is stable and all trajectories converge to it.]{
\includegraphics[width=0.3\textwidth]{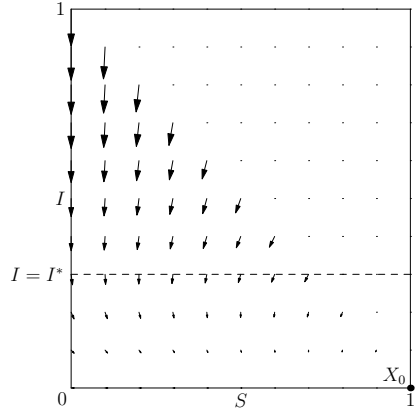}
\label{fig:vf0}
}\\
\subfloat[The case $\delta < \beta$ and $I^* > \frac{1-\frac{\delta}{\beta}}{1+\frac{\delta}{\gamma}}$. The point $X_1=\left(\frac{\delta}{\beta}, \frac{1-\frac{\delta}{\beta}}{1 + \frac{\delta}{\gamma}}\right)$ is a stable equilibrium point, similarly to the SIRS model.]{
\includegraphics[width=0.3\textwidth]{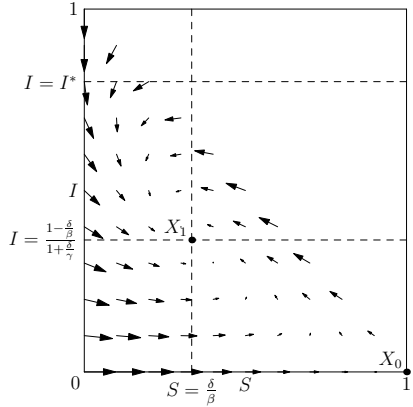}
\label{fig:X1}
}\qquad
\subfloat[The case $\delta < \beta$ and $I^* \leq \frac{1-\frac{\delta}{\beta}}{1+\frac{\delta}{\gamma}}$. The point $X_2=\left(\frac{\delta}{\beta}, I^*\right)$ is a stable equilibrium point.]{
\includegraphics[width=0.3\textwidth]{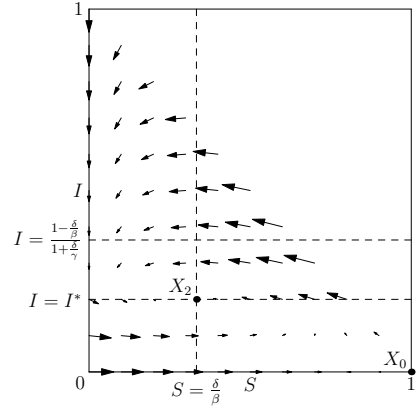}
\label{fig:X2}
}
\caption{The vector field of the system and the equilibrium points for all regions of the parameter space. At the point $(S,I)$, an arrow parallel to $(\frac{dS}{dt}, \frac{dI}{dt})$ is plotted. In cases \subref{fig:X1} and \subref{fig:X2}, the point $X_0=(1,0)$ is also an equilibrium point but it is unstable. All trajectories converge to $X_1$ or $X_2$, respectively, except those that start on the axis $I=0$, which converge to $X_0$.}
\label{fig:discont-vectorfield}
\end{figure}

\subsection{Stationary Points}\label{subsec:discont-points}

    The stationary points are found by solving for $x$ the inclusion $0\in F(x)$.

\subsubsection{Stationary points above the discontinuity line}
    There can be no stationary points in the domain $D^+$. The system becomes
    \begin{subequations}\label{eq:SystemAboveL}
    \begin{align}
      \frac{d}{dt}S& = -\beta SI - \gamma S\\
      \frac{d}{dt}I& = \beta SI - \delta I.
    \end{align}
    \end{subequations}

    From the first equation, we see that $S$ has to be zero. But then the second equation implies that $I$ also has to be zero, which is not an admissible value for $I$ as $I=0$ cannot be above the discontinuity line.

\subsubsection{Stationary points below the discontinuity line}

    We look for stationary points in the domain $D^-$. The system becomes
    \begin{subequations}\label{eq:SystemBelowL}
    \begin{align}
      \frac{d}{dt}S& = -\beta SI + \gamma (1-S-I)\\
      \frac{d}{dt}I& = \beta SI - \delta I,
    \end{align}
    \end{subequations}
    which is identical to the SIRS case (recall the discussion in the User Behavior section) except that the domain is not the whole state space, it is only $D^-$.

    This system has the solutions:
    \begin{align}
    X_0& = (S_0, I_0) = (1,0) \label{eq:X0discont}\\
    X_1& = (S_1, I_1) = \left(\frac{\delta}{\beta}, \frac{1-\frac{\delta}{\beta}}{1+\frac{\delta}{\gamma}}\right).\label{eq:X1discont}
    \end{align}
    The second solution, $X_1$, is admissible if and only if $X_1 \in D^-$, i.e.,
    \begin{equation}\label{eq:X0cond}
    \frac{\delta}{\beta} \leq 1,
    \end{equation}
    and also
    \begin{equation}\label{eq:X1cond}
      \frac{1-\frac{\delta}{\beta}}{1+\frac{\delta}{\gamma}} < I^*.
    \end{equation}

    Note that if $\frac{\delta}{\beta} = 1$, then $X_0$ and $X_1$ coincide. Also, it is not surprising that $X_1$ is the equilibrium point of the corresponding SIRS model. That is, $I^*$ does not play an explicit role in this case, as long as \eqref{eq:X1cond} holds.

\subsubsection{Stationary points on the discontinuity line}
    We look for stationary points on the discontinuity line $I=I^*$, that is, we solve the inclusion $0\in F(S, I^*)$ for $S$. The system becomes
     \begin{subequations}\label{eq:SystemOnL}
    \begin{align}
      \frac{d}{dt}S& = -\beta SI^* + [-\gamma S, \gamma (1-S-I^*)]\\
      \frac{d}{dt}I& = \beta SI^* - \delta I^*.
    \end{align}
    \end{subequations}
    Since $I^*>0$, $\frac{d}{dt}I$ is zero only when $S=\frac{\delta}{\beta}$. We then have to check if it is possible to make $\frac{d}{dt}S$ equal to zero, that is, if $0\in F(\frac{\delta}{\beta}, I^*)$. We find that it is possible when $I^*$ is such that
    \begin{equation}\label{eq:X2cond}
      I^* \leq \frac{1-\frac{\delta}{\beta}}{1+\frac{\delta}{\gamma}}.
    \end{equation}
    In that case, the stationary point is
    \begin{equation}\label{eq:X2discont}
      X_2 = (S_2, I_2) = \left(\frac{\delta}{\beta}, I^*\right).
    \end{equation}
In general, there are many combinations of $p_{SP}(I^*)$ and $p_{PS}(I^*)$ that make $\frac{d}{dt}S$ equal to zero, but there is always one with $p_{SP}(I^*)=0$. In that case, $p_{PS}(I^*) = \frac{\delta I^*}{\gamma(1-\frac{\delta}{\beta}-I^*)}$.

To summarize, $X_0$ exists always. If $\delta < \beta$, one more equilibrium point exists: $X_1$ if $I^* > \frac{1-\frac{\delta}{\beta}}{1+\frac{\delta}{\gamma}}$, or $X_2$ otherwise. In Figure~\ref{fig:discont-vectorfield} we can see these equilibrium points.

\subsection{Local Asymptotic Stability}\label{subsec:discontinuous-las}
\subsubsection{Stability of $X_0$ and $X_1$}
We show that, when $\frac{\delta}{\beta} \geq 1$, $X_0$ is asymptotically stable. When $\frac{\delta}{\beta} < 1$, $X_0$ is a saddle point, and if $X_1$ exists it is asymptotically stable.

    We now examine whether $X_0$ and $X_1$ are (locally) stable equilibrium points. The Jacobian of the system is
    \begin{equation}
      J(S,I) = \begin{pmatrix}
        -\beta I - \gamma & -\beta S - \gamma\\
        \beta I & \beta S - \delta
      \end{pmatrix}
    \end{equation}

    We evaluate the Jacobian at the point $X_0$:
    \begin{equation}
      J(X_0) = J(0,1) = \begin{pmatrix}
        - \gamma & -\beta - \gamma\\
        0 & \beta - \delta
      \end{pmatrix}
    \end{equation}

    The eigenvalues of $J(X_0)$ are $-\gamma$ and $\beta - \delta$. So, $X_0$ is stable if and only if $\beta < \delta$, in which case note that $X_1$ does not exist.

    We evaluate the Jacobian at the point $X_1$:
    \begin{equation}
    \begin{split}
      J(X_1)& = J\left(\frac{\delta}{\beta}, \frac{1-\frac{\delta}{\beta}}{1+\frac{\delta}{\gamma}}\right) \\
      &= \begin{pmatrix}
        -\frac{\beta + \gamma}{1+\frac{\delta}{\gamma}} & -\delta - \gamma\\
        \frac{\beta - \delta }{1+\frac{\delta}{\gamma}} & 0
      \end{pmatrix}
    \end{split}
    \end{equation}

    The eigenvalues of $J(X_1)$ are $\frac{a_{11}\pm \sqrt{a_{11}^2 + 4a_{12}a_{21}}}{2}$, where $a_{ij}$ are the elements of $J(X_1)$ ($a_{22}=0$). Since $a_{11}<0$, the smallest eigenvalue is always negative. The largest one is negative if and only if $a_{12}a_{21}<0 \Equiv \beta > \delta$. So $X_1$ is stable whenever it exists.

     If we evaluate the square root $\sqrt{a_{11}^2 + 4a_{12}a_{21}}$ at the point $\beta = \gamma(1+\frac{\gamma}{\delta})^2 - \frac{\gamma}{2}$, we see that its argument can also take negative values. Since the eigenvalues are a continuous function of $\beta$, they will have an imaginary part for $\beta$ close to $\gamma(1+\frac{\gamma}{\delta})^2 - \frac{\gamma}{2}$, which means that the trajectories \emph{spiral} towards $X_1$.

\subsubsection{Stability of $X_2$}

    To show that the stationary point on the discontinuity line is asymptotically stable, we will use Theorem~\ref{thm:stability} below \cite[\S19, Theorem 3]{filippov-book}. To use this theorem we transform the system so that the line of discontinuity is the horizontal axis, the stationary point is $(0,0)$, and the trajectories have a clockwise direction for increasing $t$.

    We set $x=\frac{\delta}{\beta} - S$ and $y=I-I^*$. The domains $D,D^-,D^+$ become $G = \{(x,y) | x \leq \frac{\delta}{\beta}, y \geq -I^*, y-x \leq 1-I^*-\frac{\delta}{\beta}\}$,  $G^- = G\cap\{(x,y) | y<0\}$, and $G^+ = G\cap\{(x,y) | y>0\}$. Then, the system can be written as
    \begin{subequations}
    \begin{align}
      \frac{dx}{dt}& = P^-(x,y) = -\beta xy - (\beta I^* + \gamma)x + (\gamma + \delta)y -\gamma(1-I^*) + \delta(I^* + \frac{\gamma}{\beta})\\
      \frac{dy}{dt}& = Q^-(x,y) = -\beta x(y+I^*)
    \end{align}
    \end{subequations}
    for $(x,y) \in G^-$, and
    \begin{subequations}
    \begin{align}
      \frac{dx}{dt}& = P^+(x,y) = -\beta xy - (\beta I^* + \gamma)x + \delta y + \delta(I^* + \frac{\gamma}{\beta})\\
      \frac{dy}{dt}& = Q^+(x,y) = -\beta x(y+I^*)
    \end{align}
    \end{subequations}
    for $(x,y) \in G^+$.

    The partial derivatives of $P^\pm$, that is, of $P^+$ and of $P^-$, are denoted by $P^\pm_x, P^\pm_{xx}, P^\pm_y$ etc., and similarly for $Q^\pm$. We define two quantities $A^\pm$ in terms of the functions $P^\pm$, $Q^\pm$ and their derivatives at the point $(0,0)$:
    \begin{equation}
        A^\pm = \frac{2}{3}\left(\frac{P^\pm_x + Q^\pm_y}{P^\pm} - \frac{Q^\pm_{xx}}{2Q^\pm_x}\right).
    \end{equation}

    \begin{theorem}\label{thm:stability}
      Let the conditions
      \begin{gather}
        Q^- = Q^+ = 0, P^- < 0, P^+ > 0\\
        Q_x^- < 0, Q_x^+ < 0
      \end{gather}
      be fulfilled at the point $(0,0)$. Then, $A^+ - A^- < 0$ implies that the zero solution is asymptotically stable, whereas $A^+ - A^- > 0$ implies that the zero solution is unstable.
    \end{theorem}

    All the conditions of Theorem~\eqref{thm:stability} are satisfied in our case, together with $A^+ - A^- < 0$. The condition $P^- < 0$ is equivalent to \eqref{eq:X2cond}, i.e., the condition on $I^*$ that causes the stationary point to be on the line of discontinuity. All the other conditions are straightforward to verify. For example, to prove that $A^+ - A^- < 0$ we can quickly establish that $A^+ < 0$ and $A^- > 0$, again using \eqref{eq:X2cond}.

    Therefore, the stationary point $(S,I) = (\frac{\delta}{\beta}, I^*)$ is asymptotically stable.

\subsection{Domains of Attraction}\label{subsec:discontinuous-domains}

From Theorem 6, \S 13 \cite{filippov-book} we know that for autonomous systems on the plane, it holds that if a half trajectory $T^+$ is bounded, then its $\omega$-limit set $\Omega(T^+)$ contains either a stationary point or a closed trajectory. Recall that the $\omega$-limit set of a half trajectory $T^+ (x=\phi(t), t_0\leq t < \infty)$ is the set of all points $q$ for which there exists a sequence $t_1, t_2, \ldots$ tending to $\infty$ such that $\phi(t_i)\To q$ as $i\To\infty$.

In this section, we show that there are no solutions that are closed trajectories. So we can conclude that all system trajectories converge to equilibrium points. When there is more than one equilibrium point, we show which trajectories converge to which point.

The main result is that for any half trajectory $T^+$, its $\omega$-limit set $\Omega(T)$ can only contain equilibrium points, that is, $X_0 = (1,0)$, $X_1 = (S_1, I_1) = \left(\frac{\delta}{\beta}, \frac{1-\frac{\delta}{\beta}}{1+\frac{\delta}{\gamma}}\right)$, or $X_2 = (\frac{\delta}{\beta}, I^*)$.

We will find the following two functions useful:
\begin{align}
  E(S,I)& = S-S_1\ln(S) + I + \frac{\gamma}{\beta}\ln(I), \quad (S,I)\in D^+\\
  M(S,I)& = S -(S_1+\frac{\gamma}{\beta})\ln(S+\frac{\gamma}{\beta}) + I -I_1\ln(I), \quad (S,I)\in D^-.
\end{align}
It holds that $E(S,I)$ is constant on trajectories in the area $D^+$, and $M(S,I)$ is decreasing along trajectories in the area $D^-$. Indeed, with some calculations it can be shown that
\begin{align}
  \frac{d}{dt} E(S,I)& = \frac{\partial E}{\partial S}\frac{dS}{dt} + \frac{\partial E}{\partial I}\frac{dI}{dt} = 0\\
  \frac{d}{dt} M(S,I)& = \frac{\partial M}{\partial S}\frac{dS}{dt} + \frac{\partial M}{\partial I}\frac{dI}{dt} = -\frac{(\beta S - \delta)^2}{\beta S + \gamma} \frac{1+\frac{\gamma}{\beta}}{1+\frac{\delta}{\gamma}} \leq 0.
  \end{align}

First of all, we prove that a trajectory converges to $X_0 = (1,0)$ if and only if it starts on the line $I=0$: If it starts on the line, then $\frac{d}{dt}I$ is zero, so $I$ stays equal to 0, and the trajectory stays on the line. And $\frac{d}{dt}S$ is positive always except on $X_0$, so the trajectory converges to $X_0$.

If the trajectory starts at a point $(S_0, I_0), I_0>0$, then let $M(S_0, I_0)=M_0$. We can see that, for any $S$ it holds that $\lim_{I\rightarrow 0} M(S,I) = \infty$. So, if the trajectory comes close enough to the line $I=0$, the function $M(S,I)$ will have to increase above $M_0$, which is a contradiction. Therefore, the trajectory cannot converge to $X_0 = (1,0)$.

From now on, we assume that on all points of a trajectory it holds that $I>0$.

Assume that there exists a half trajectory $T^+$ whose limit set $\Omega(T)$ contains a closed trajectory $\Gamma$. By successively eliminating properties of such a trajectory, we will prove that it cannot exist. Note that Lemma~\ref{lem:lemma1} below is trivial if $(\frac{\delta}{\beta}, I^*)$ is an equilibrium point.

\begin{lemma}\label{lem:lemma1}
  The point $(\frac{\delta}{\beta}, I^*)$ cannot be on $\Gamma$.
\end{lemma}
\begin{IEEEproof}

If $(S, I)=(\frac{\delta}{\beta}, I^*)$ is on $\Gamma$ (say, at time $t_1$), then, first of all, $0 \notin F(\frac{\delta}{\beta}, I^*)$, because otherwise the point $(\frac{\delta}{\beta}, I^*)$ would be an equilibrium point, so it could not be part of a closed trajectory. Since $0 \notin F(\frac{\delta}{\beta}, I^*)$, the point $X_1$ is an equilibrium and it is distinct from $(\frac{\delta}{\beta}, I^*)$.

We will now reach a contradiction by proving that $\Gamma$ cannot approach $(S, I)=(\frac{\delta}{\beta}, I^*)$ for $t>t_1$, thus $\Gamma$ cannot be a closed trajectory. If $(\frac{\delta}{\beta}, I^*)\in \Gamma$, then the trajectory would have to exit the line $L$ immediately after passing through it (otherwise right uniqueness would be violated on the points of $L$ contained in $\Gamma$).

Define a region $\Delta_\epsilon \subseteq D^-$ around $(\frac{\delta}{\beta}, I^*)$ that includes all points where the $M$ function has values at least equal to $M(\frac{\delta}{\beta}, I^*) - \epsilon$, for some small enough $\epsilon>0$ such that $\Delta_\epsilon$ does not include $X_1$. Since $M$ is continuous and $X_1$ is distinct from $(\frac{\delta}{\beta}, I^*)$, $\Delta_\epsilon$ is well defined.

$\Gamma$, being a closed trajectory, has to encircle an equilibrium point (Thm. 7, \S 13 \cite{filippov-book}). So, it would have to exit $\Delta_\epsilon$ and go around the point $X_1$. But then, as $M$ decreases along trajectories, $\Gamma$ cannot reenter $\Delta_\epsilon$ before exiting $D^-$. It can only exit by crossing the line $L$ somewhere on the interval $((\frac{\delta}{\beta}, I^*), (1-I^*, I^*)]$. However, at $(\frac{\delta}{\beta}, I^*)$ the function $M(S,I^*)$ attains a minimum over $S$, so $\Gamma$ cannot exit $D^-$ because it cannot approach $L$.
\end{IEEEproof}

If $(S, I)=(\frac{\delta}{\beta}, I^*)$ is not on $\Gamma$, then on $\Gamma$ there holds right uniqueness. Also, $\Omega(\Gamma)=\Gamma$.

We will continue by proving that $\Gamma$ cannot have more or fewer than two intersection points with $L$.

\begin{lemma}
A closed trajectory $\Gamma$ that does not pass through the point $(\frac{\delta}{\beta}, I^*)$ cannot have either more than two or fewer than two intersection points with the discontinuity line $L$. If it has two intersection points, they cannot be on the same side of $(\frac{\delta}{\beta}, I^*)$.
\end{lemma}
\begin{IEEEproof}

Denote by $\Gamma\cap L = \{l_1, l_2, l_3, \ldots\}$ the common points of $\Gamma$ with $L$, and $t_1, t_2, t_3, \ldots$ the corresponding times.

$\Gamma\cap L$ cannot be empty, because $\Gamma$ cannot be completely contained within the area $D^-$, because the function $M$ is decreasing in $D^-$, nor within the area $D^+$, because $\Gamma$ has to encircle an equilibrium point, but there is no equilibrium point in $D^+$.

$\Gamma\cap L$ cannot be a singleton set. If there is only one point in $\Gamma\cap L$, say $l_1$, then $\Gamma$ has to be in $D^-$ (except for $l_1$) because it has to encircle $X_1$. Then, $\Gamma$ has to exit $L$ immediately, otherwise it would have more than one common points with $L$. If $\Gamma$ exits an $\epsilon$-neighborhood of $l_1$, then, using the function $M(S,I)$ we can show that $\Gamma$ cannot return in an appropriate $\delta$-neighborhood of $l_1$, so $\Gamma$ cannot be closed.

Let there be 3 or more distinct points in $\Gamma\cap L$. At least two of these points are on the same side of the point $(\frac{\delta}{\beta}, I^*)$, assume the side on the right ($S > \frac{\delta}{\beta}$). Call them $l_i$ and $l_j$, and their corresponding times $t_i$ and $t_j$. Assume without loss of generality that $l_i$ is the one closer to $(\frac{\delta}{\beta}, I^*)$. Since $l_i$ is distinct from $(\frac{\delta}{\beta}, I^*)$, there is at least one more point on $L$ between $(\frac{\delta}{\beta}, I^*)$ and $l_i$. Call that point $\alpha$, and consider the line segment $LS$ from $\alpha$ to $(1-I^*, I^*)$. By construction, both $l_i$ and $l_j$ are on $LS$. The segment $LS$ is a \emph{transversal}: it is intersected by trajectories only in one direction, as $\frac{d}{dt}I>0$ for $S > \frac{\delta}{\beta}$. Also, right uniqueness holds on the points of $LS$.

By Lemma 3 \S13 in \cite{filippov-book}, for a trajectory $T$ the set $\Omega(T)$ can intersect the transversal $LS$ at not more than one point. So, since  $\Gamma = \Omega(\Gamma)$, the set $\Gamma\cap LS$ cannot contain more than one point, so we have a contradiction. We reach a similar contradiction if we assume that $l_i$ and $l_j$ are to the left of $(\frac{\delta}{\beta}, I^*)$.
\end{IEEEproof}

\begin{lemma}
  A closed trajectory $\Gamma$ cannot intersect the discontinuity line $L$ on exactly two points that are on opposite sides of the point $(\frac{\delta}{\beta}, I^*)$.
\end{lemma}
\begin{IEEEproof}

Call $A=(S_A, I^*)$ the point in $\Gamma\cap L$ with $S_A<\frac{\delta}{\beta}$, and call $B=(S_B, I^*)$ the one with $S_B>\frac{\delta}{\beta}$.

Let $\Gamma$ be parameterized by $\phi(t) = (x(t), y(t)), t\in [0,T]$; also $\phi(0)=\phi(T)$. The function $\phi(t)$ is a solution of the differential inclusion, that is, $\dot{\phi}(t) = (\dot{x}(t), \dot{y}(t)) \in F(\phi(t)), t\in [0,T]$. Let $t_A, t_B\in [0,T]$ be such that $A=\phi(t_A)$ and $B=\phi(t_B)$. Let $\alpha_A, \alpha_B \in [0,1]$ be such that $\dot{x}(t_A) = -\beta xy -\gamma x + \alpha_A\gamma(1-y)$ and $\dot{x}(t_B) = -\beta xy -\gamma x + \alpha_B\gamma(1-y)$

Define the functions $P(x,y)$ and $Q(x,y), (x,y)\in D\setminus\{y, y>0\}$:
\begin{equation}
  P(x,y) = -\frac{1}{y}\dot{y} = -\frac{1}{y}(\beta xy - \delta y) = \delta - \beta x
\end{equation}

\begin{equation}
  Q(x,y) = \frac{1}{y}\dot{x} = \begin{cases}
    \frac{1}{y}(-\beta xy + \gamma(1-x-y)), & y<I^*\\
    \frac{1}{y}(-\beta xy -\gamma x + \alpha_A\gamma(1-y)), & x\leq\frac{\delta}{\beta}, y=I^*\\
    \frac{1}{y}(-\beta xy -\gamma x + \alpha_B\gamma(1-y)), & x>\frac{\delta}{\beta}, y=I^*\\
    \frac{1}{y}(-\beta xy -\gamma x), & y>I^*
  \end{cases}
\end{equation}

We compute the integral $\oint_\Gamma Pdx + Qdy$ in two ways.

For the first computation, we use the parametrization $\phi(t) = (x(t), y(t))$ of $\Gamma$, so $dx = \dot{x}dt$ and $dy = \dot{y}dt$. The result is zero:
\begin{equation}\label{eq:firstcomputation}
  \oint_\Gamma Pdx + Qdy = \int_0^T -\frac{1}{y}\dot{y}\dot{x}dt + \frac{1}{y}\dot{x}\dot{y}dt = 0.
\end{equation}

For the second computation, we split $Q(x,y)$ into two functions, one continuous and one discontinuous, so that $Q(x,y) = Q_c(x,y)+Q_d(x,y)$.
\begin{equation}
  Q_c(x,y) = \frac{1}{y}(-\beta xy -\gamma x)
\end{equation}
\begin{equation}
  Q_d(x,y) = \begin{cases}
    \frac{1}{y}\gamma(1-y), & y<I^*\\
    \frac{1}{y}\alpha_A\gamma(1-y), & x\leq\frac{\delta}{\beta}, y=I^*\\
    \frac{1}{y}\alpha_B\gamma(1-y), & x>\frac{\delta}{\beta}, y=I^*\\
    0, & y>I^*
  \end{cases}
\end{equation}
So now the original integral can be split into two:
$\oint_\Gamma Pdx + Qdy = \oint_\Gamma Pdx + (Q_c+Q_d)dy = \oint_\Gamma Pdx + Q_cdy + \oint_\Gamma Q_d dy$. We use Green's theorem to compute the first integral.
\begin{equation}\label{eq:secondcomputationpart1}
  \oint_\Gamma Pdx + Q_cdy = \iint_\Gamma \frac{\partial Q_c}{\partial x} - \frac{\partial P}{\partial y} dxdy = \iint_\Gamma -\beta -\frac{\gamma}{y} dxdy < 0.
\end{equation}

For the second integral $\oint_\Gamma Q_d dy$, we define the function
\begin{equation}
  Q_d^{ext}(x,y) = \frac{1}{y}\gamma(1-y), \qquad (x,y)\in D\setminus\{y, y>0\}
\end{equation}
and the curves $\Gamma_1$ and $\Gamma_2$: The curve $\Gamma_1$ is the trajectory $\Gamma$ restricted to $y\leq I^*$. The direction of $\Gamma_1$ is from $A$ to $B$. The curve $\Gamma_2$ is the line segment of $L$ joining $B$ and $A$, with direction from $B$ to $A$.

Observe that
\begin{equation}\label{eq:secondcomputationpart2}
  \oint_\Gamma Q_d dy = \oint_{\Gamma_1\cup\Gamma_2} Q_d^{ext} dy = \iint_{\Gamma_1\cup\Gamma_2} \frac{\partial Q_d^{ext}}{\partial x}dxdy = 0,
\end{equation}
where the first equality follows from $Q_d \equiv Q_d^{ext}$ on $\Gamma_1$ and $dy=0$ on $\Gamma_2$, whereas the last equality follows from Green's theorem, because $Q_d^{ext}$ is continuously differentiable.

We see that the result of \eqref{eq:firstcomputation} contradicts the result of \eqref{eq:secondcomputationpart1} and \eqref{eq:secondcomputationpart2}. So, the trajectory $\Gamma$ with the assumed properties cannot exist.

\end{IEEEproof}

From the previous lemmata, we conclude that there can be no closed trajectory $\Gamma$. Therefore, all trajectories have to converge to equilibrium points.

\subsection{Conclusion}
\begin{figure}[htbp]
  \centering
  \includegraphics[height = 6cm]{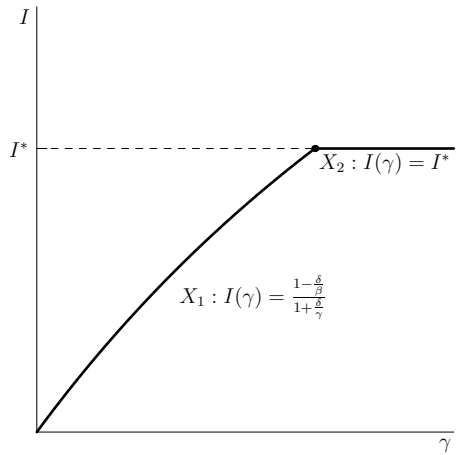}\\
  \caption{The total fraction of Infected as a function of $\gamma$.}\label{fig:discont-IofGamma}
\end{figure}

In Figure~\ref{fig:discont-IofGamma} we see that the total fraction $I=\frac{1-\frac{\delta}{\beta}}{1+\frac{\delta}{\gamma}}$ of Infected at the system equilibrium increases with the update rate $\gamma$, until $I$ becomes equal to the threshold $I^*$. The reason for this increase is that, when the equilibrium value of $I$ is below $I^*$, the trajectories will eventually be completely contained in the domain $D^-$ (below $I^*$). In this domain, at each time a Protected is informed about the value of $I$, he will choose to become Susceptible, thus fueling the infection. In parallel, no Susceptible will choose to become Protected. The larger the value of $\gamma$, the shorter time a user will spend being Protected, thus the smaller the fraction of Protected. However, a smaller fraction of Protected implies a larger fraction of Infected, as the fraction of Susceptible at equilibrium is necessarily $\frac{\delta}{\beta}$, i.e., it is independent of $\gamma$.

When the quantity $\frac{1-\frac{\delta}{\beta}}{1+\frac{\delta}{\gamma}}$ exceeds $I^*$, the equilibrium value of $I$ is limited to $I^*$; further increases of $\gamma$ have no effect. The explanation is that, as soon as the instantaneous value of $I$ exceeds $I^*$, Susceptible users switch to Protected, and Protected users stay Protected, thus bringing the infection level below $I^*$. However, there is no equilibrium point for the system in the domain $D^-$, so the only possible equilibrium value of $I$ is $I^*$. For $I=I^*$ there are in general many combinations of $p_{SP}(I^*)$ and $p_{PS}(I^*)$ that lead to an equilibrium, including one with $p_{SP}(I^*)=0$ and $p_{PS}(I^*)>0$. That combination means that no Susceptible users become Protected, but some Protected become Susceptible. Other combinations with both $p_{SP}(I^*)>0$ and $p_{PS}(I^*)>0$ would be harder to justify, as they imply that at the same value of $I^*$ users would switch from Susceptible to Protected and back.

A side conclusion concerns the interaction of $\gamma$ with $I^*$: We have seen that increasing $\gamma$ will increase the equilibrium value of $I$, but $I$'s maximum value will be limited to $I^*$, so if $I^*$ is low, the effect of increasing $\gamma$ is not severe.

\section{The Users are not Strictly Rational}\label{sec:continuous}
The user behavior functions $p_{SP}(I)$ and $p_{PS}(I)$ are continuously differentiable, and we require that $\frac{d}{dI}p_{SP}(I) > 0$ and $\frac{d}{dI}p_{SP}(I) < 0$. Other than that, the two functions are arbitrary.

 \subsection{Stationary Points}
 The equilibrium points of the system are found by solving for $x$ the equation $F(x) = 0$:
    \begin{subequations}\label{eq:dup-mastersystem}
    \begin{align}
      \frac{d}{dt}S = 0& = -\beta SI -\gamma Sp_{SP}(I) + \gamma (1-S-I)p_{PS}(I) \label{eq:dup-mastersystem1}\\
      \frac{d}{dt}I = 0& = \beta SI - \delta I \label{eq:dup-mastersystem2}
    \end{align}
    \end{subequations}
From \eqref{eq:dup-mastersystem2} we see that either $I=0$ or $S=\frac{\delta}{\beta}$.
\begin{itemize}
  \item Equilibrium point $X_0$\\
  Substituting $I=0$ into \eqref{eq:dup-mastersystem1}, we have that $X_0 = (S_0, I_0) = \left(\frac{p_{PS}(0)}{p_{SP}(0) + p_{PS}(0)}, 0\right)$. These values of $(S_0, I_0)$ are always admissible as they are always non-negative and at most equal to 1.

  Recalling the meaning of $p_{PS}(0)$ and $p_{SP}(0)$, we can reasonably expect that $p_{PS}(0) = 1$ and $p_{SP}(0) = 0$: Protected have no reason to remain Protected, and Susceptible have no reason to become Protected, when there is no infection in the network. In this case, $X_0$ is the point $(1,0)$.

  \item Equilibrium point $X_1$\\
  Substituting $S=\frac{\delta}{\beta}$ into \eqref{eq:dup-mastersystem1}, we see that $I$ has to satisfy
  \begin{equation}
      g(I) \equiv -\delta I -\frac{\gamma\delta}{\beta}p_{SP}(I) + \gamma \left(1-\frac{\delta}{\beta}-I\right)p_{PS}(I) = 0.
  \end{equation}
  To solve $g(I)=0$ for $I$ we need to know the two response functions $p_{SP}(I)$ and $p_{PS}(I)$. But even without knowing them, we can still prove that $g(I)=0$ has a unique solution for $I\in [0,1]$ under the condition that
  \begin{equation}\label{eq:dup-condition}
      \frac{\delta}{\beta} \leq \frac{p_{PS}(0)}{p_{SP}(0) + p_{PS}(0)}.
  \end{equation}
  We first show that $g(I)$ monotonically decreases in the interval $[0,1]$, and then we show that, under the condition \eqref{eq:dup-condition}, $g(0)g(1)\leq 0$. We can then conclude that there is exactly one solution of $g(I)=0$ in the interval $[0,1]$.
    \begin{equation}
      \frac{d}{dI}g(I) = -\delta - \frac{\gamma\delta}{\beta}\frac{d}{dI}p_{SP}(I) - \gamma p_{PS}(I)+ \gamma \left(1-\frac{\delta}{\beta}-I\right)\frac{d}{dI}p_{PS}(I)
    \end{equation}
    And, as $\frac{d}{dI}p_{SP}(I) > 0$ and $\frac{d}{dI}p_{SP}(I) < 0$, we can see that
    \begin{equation}
      \frac{dg(I)}{dI} < 0, \forall \beta,\gamma,\delta>0,
    \end{equation}
    so $g(I)$ monotonically decreases.

    Under the condition~\eqref{eq:dup-condition}, $g(0)$ is non-negative:
    \begin{equation}
      g(0) = -\frac{\gamma\delta}{\beta}p_{SP}(0) + \gamma \left(1-\frac{\delta}{\beta}\right)p_{PS}(0)
    \end{equation}

    \begin{equation}
      g(0) \geq 0 \Equiv \frac{\delta}{\beta} \leq \frac{p_{PS}(0)}{p_{SP}(0) + p_{PS}(0)},
    \end{equation}
    which is exactly condition \eqref{eq:dup-condition}.

    And now we prove that $g(1)$ is always negative.
    \begin{equation}
      g(1) = -\delta -\frac{\gamma\delta}{\beta}p_{SP}(1) - \gamma\frac{\delta}{\beta}p_{PS}(1)
    \end{equation}
    Therefore:
    \begin{equation}
      g(1) < 0, \forall \beta,\gamma,\delta>0.
    \end{equation}

    Denoting by $I_1$ the solution of $g(I)=0$, we can now conclude that $X_1 = (S_1, I_1) = (\frac{\delta}{\beta}, I_1)$ is uniquely determined under \eqref{eq:dup-condition}. The values $S_1, I_1$ are admissible as they are both between 0 and 1. Note that if \eqref{eq:dup-condition} does not hold then both $g(0)<0$ and $g(1)<0$, so the monotonicity of $g$ in $[0,1]$ implies that $X_1$ does not exist. Consequently, \eqref{eq:dup-condition} is both necessary and sufficient for the existence of $X_1$.
  \end{itemize}
  \subsection{Local Asymptotic Stability}
  To examine the (local) stability of the equilibrium points $X_0$ and $X_1$ we compute the Jacobian of the system \eqref{eq:dup-mastersystem} and evaluate it at these two points.

       \begin{equation}
      J(S,I) = \begin{pmatrix}
        j_{11} & j_{12}\\
        j_{21} & j_{22}
      \end{pmatrix}
    \end{equation}
where
\begin{subequations}
\begin{align}
  j_{11} =& -\beta I - \gamma(p_{SP}(I) + p_{PS}(I)) \\
  j_{12} =& -\beta S - \gamma S \frac{d}{dI}p_{SP}(I) - \gamma p_{PS}(I) + \gamma(1-S-I)\frac{d}{dI}p_{SP}(I)\\
  j_{21} =& \beta I\\
  j_{22} =& \beta S - \delta
\end{align}
\end{subequations}
Observe that both $j_{11}$ and $j_{12}$ are negative for all $S$ and $I$: recall our assumption that $\frac{d p_{SP}(I)}{dI} > 0$ and $\frac{d p_{SP}(I)}{dI} < 0$.

For the case of $X_0$, the Jacobian is
\begin{equation}
      J(X_0) = \begin{pmatrix}
        -\gamma(p_{SP}(0) + p_{PS}(0)) & J(X_0)_{12}\\
        0 & \beta \frac{p_{PS}(0)}{p_{SP}(0) + p_{PS}(0)} - \delta
      \end{pmatrix}
\end{equation}
where the value of $J(X_0)_{12}$ is irrelevant for the calculation of the eigenvalues. The two eigenvalues of $J(X_0)$ are the two diagonal entries:
\begin{align}
  \lambda_1^0 = & -\gamma(p_{SP}(0) + p_{PS}(0))\\
  \lambda_2^0 = & \beta \frac{p_{PS}(0)}{p_{SP}(0) + p_{PS}(0)} - \delta
\end{align}

The first eigenvalue, $\lambda_1^0$, is always negative except in the trivial case where $p_{SP}(0) = p_{PS}(0) = 0$. The second one is negative if and only if
\begin{equation}
    \frac{\delta}{\beta} > \frac{p_{PS}(0)}{p_{SP}(0) + p_{PS}(0)},
\end{equation}
which is exactly the opposite of condition \eqref{eq:dup-condition}. So, $X_0$ is stable when $X_1$ does not exist.

For the case of $X_1$ the Jacobian is
    \begin{equation}
      J(X_1)= \begin{pmatrix}
         J(X_1)_{11}& J(X_1)_{12}\\
         \beta I_1 & 0
      \end{pmatrix}
    \end{equation}
The eigenvalues have the following form:
\begin{equation}
  \lambda_1^1, \lambda_2^1 = \frac{1}{2}\left(J(X_1)_{11}\pm \sqrt{J(X_1)_{11}^2 + 4\beta I_1 J(X_1)_{12}} \right)
\end{equation}

They are both negative, since $J(X_1)_{11}$ and $J(X_1)_{12}$ are negative as mentioned before. So, $X_1$ is stable whenever it exists.

  \subsection{Domains of Attraction}
  Since the system is two-dimensional and $F$ is continuously differentiable, we can use Dulac's criterion to show that the system can have no periodic trajectory.
  \begin{theorem}[Dulac's criterion]
    Let $A$ be a simply connected domain. If there exists a continuously differentiable function $h: A \To \mathbb{R}$ such that $\nabla\cdot(hF)$ is continuous and non-zero on $A$, then no periodic trajectory can lie entirely in $A$.
  \end{theorem}

  In our case, the domain $A$ is the state space excluding the line $I=0$. Note that there can be no periodic trajectory that passes from a point with $I=0$. We select as function $h$ the function $h(S,I)=\frac{1}{I}$. We compute $\nabla\cdot(hF)$ to be
  \begin{equation}
    \nabla\cdot(hF) = -\beta -\gamma\frac{p_{SP}(I)}{I} - \gamma\frac{p_{PS}(I)}{I} < 0, \forall (S,I)\in A,
  \end{equation}
  which is continuous and non-zero in $A$. Then, from Dulac's criterion, no periodic trajectory lies entirely in $A$, and, consequently, the system has no periodic trajectory at all. From the Poincar\'e-Bendixson theorem, the system can only converge to a periodic trajectory or an equilibrium point; so, we can conclude that every trajectory must converge to an equilibrium point, that is, either to $X_0$ or $X_1$.

  More precisely, when \eqref{eq:dup-condition} does not hold, only $X_0$ exists so all trajectories converge to $X_0$. When \eqref{eq:dup-condition} holds, both $X_0$ and $X_1$ exist, and $X_0$ is a saddle point: Trajectories starting on the line $I=0$ approach $X_0$ along the line $I=0$, whereas all other trajectories converge to $X_1$.
  Indeed, if $I(0)>0$, then the corresponding trajectory will have $I(t)>0, \forall t>0$. The reason is that if $I(t_0)=0$ for some finite $t_0>0$, then the uniqueness of solutions would be violated at $(S(t_0),I(t_0))$, because it would be a common point with the trajectories that approach $X_0$ along the line $I=0$. If $t_0 =\infty$, i.e., the trajectory with $I(0)>0$ converges asymptotically to $X_0$ while keeping $I(t)>0$, then close enough to $X_0$ we reach a contradiction as $\frac{d}{dt}I$ will become positive (see \eqref{eq:dup-condition} and \eqref{eq:dup-mastersystem2}).

  \subsection{Conclusion}
The equilibrium point $X_0$ is unaffected by $\gamma$. We show now that, at $X_1=\left(\frac{\delta}{\beta}, I_1\right)$, the equilibrium level of the Infected increases with $\gamma$. To this end, we take the derivative $\frac{dI_1}{d\gamma}$ and we see that is always positive.

We know that $I_1$ satisfies $g(I_1)=0$, i.e.,
\begin{equation}\label{eq:gIequals0}
  -\delta I_1 -\frac{\gamma\delta}{\beta}p_{SP}(I_1) + \gamma (1-\frac{\delta}{\beta}-I_1)p_{PS}(I_1)=0
\end{equation}

Differentiating with respect to $\gamma$, then using the chain rule, and finally collecting terms, we have
\begin{equation}
\begin{split}
\frac{dI_1}{d\gamma}\left(-\delta  - \frac{\gamma\delta}{\beta}\frac{dp_{SP}(I_1)}{dI_1} - \gamma p_{PS}(I_1) + \gamma (1-\frac{\delta}{\beta}-I_1)\frac{dp_{PS}(I_1)}{dI_1}\right) \\
= \frac{\delta}{\beta}p_{SP}(I_1) - (1-\frac{\delta}{\beta}-I_1)p_{PS}(I_1).
\end{split}
\end{equation}
The term in the parenthesis on the left-hand side is negative, and so is the right-hand side. Therefore, $\frac{dI_1}{d\gamma}$ is positive.

The negativity of the left-hand side parenthesis is deduced from $\frac{dp_{SP}(I_1)}{dI_1}>0$ and $\frac{dp_{PS}(I_1)}{dI_1}<0$.
The negativity of $\frac{\delta}{\beta}p_{SP}(I_1) - (1-\frac{\delta}{\beta}-I_1)p_{PS}(I_1)$ is deduced from \eqref{eq:gIequals0}: $\frac{\delta}{\beta}p_{SP}(I_1) - (1-\frac{\delta}{\beta}-I_1)p_{PS}(I_1) = - \frac{\delta}{\gamma} I_1 < 0$.

\section{The Users have Different Behavior Functions}\label{sec:multiple}

    So far, we have assumed that all users behave in the same way, which might be unrealistic in practice. In this section, we model the case when users are split into two classes, each with a different threshold behavior function. Note that we choose to have two classes to keep the presentation simple, but we believe that our results carry over to multiple user classes.

    A fraction $a_c$ of users are in class $c=1,2$, and $a_1+a_2=1$. The fractions of Susceptible, Infected, and Protected in class $c$ are denoted by $S^c, I^c, P^c$, and it holds that $S^c + I^c + P^c = a_c$. Users do not change classes, so a user in $S^1$ will move, if infected, to $I^1$ and then to $P^1$. The total fraction of Susceptible users is denoted by $S(=S^1+S^2)$, and similarly for Infected, $I(=I^1+I^2)$, and for Protected $P(=P^1+P^2)$. Susceptible users can be infected by an Infected user \emph{of any class}, not just by an Infected of their own class.

    Users in different classes differ in their response to received alerts, i.e., the functions $p_{SP}(I)$ and $p_{PS}(I)$ become $p_{SP}^c(I)$ and $p_{PS}^c(I)$ for users in class $c$. Note that the functions depend on $I$, not on $I^c$. We will study discontinuous best response functions with a different threshold $I^{*c}$ for each class. Without loss of generality, we require $I^{*1}<I^{*2}$.
    \begin{equation}
    p_{SP}^c(I) =
       \begin{cases}
        0       & I < I^{*c}\\
        [0,1]   & I = I^{*c}\\
        1       & I > I^{*c}\\
       \end{cases}\qquad
           p_{PS}^c(I) =
       \begin{cases}
        1       & I < I^{*c}\\
        [0,1]   & I = I^{*c}\\
        0       & I > I^{*c}\\
       \end{cases}
    \end{equation}

    The first class of users that we will model are users with a low threshold $I^{*1}=0.1$, who we call Responsible. Because of their low threshold, these users become Protected easily, but they do not easily switch from Protected to Susceptible. We call them Responsible because the way they behave helps reduce the infection. The second class of users, who we call Selfish, have a high threshold $I^{*1}=0.9$. This means that they hardly ever decide to switch from Susceptible to Protected, whereas they almost always decide to leave the Protected state.

    In the next section, we simulate the system on human mobility traces, and we confirm our previous conclusion that the equilibrium level of infection increases with the update rate $\gamma$.
    
\section{Simulations on Mobility Traces}\label{sec:validation}

We validate our conclusions using simulations on human mobility traces. The traces that we use are Bluetooth contacts among 41 devices given to participants in a conference \cite{cambridge-haggle-2006-09-15}. The traces were collected over a period of approximately 72 hours.

The contact rate $\beta$ is determined by the traces. Actually, $\beta$ is a function of time $\beta(t)$, since the number of contacts per time unit fluctuates depending on the time of day. We want to establish whether the fraction of Infected indeed increases for larger values of the update rate $\gamma$. For the simulations that follow, we set $\delta = (6hr)^{-1}$, and we plot the system trajectories on the $S-I$ plane (average of 30 simulations) for three different values of $\gamma$, $(1hr)^{-1}$, $(6hr)^{-1}$, and $(24hr)^{-1}$. The initial conditions for all simulations were 1 Infected and 40 Susceptible. In the case of two user classes, the initially Infected user is of class 2 (Selfish). Each simulation runs until either there are no Infected, or the end of the traces is reached.

\begin{figure}[htbp]
  \centering\includegraphics[height = 6cm]{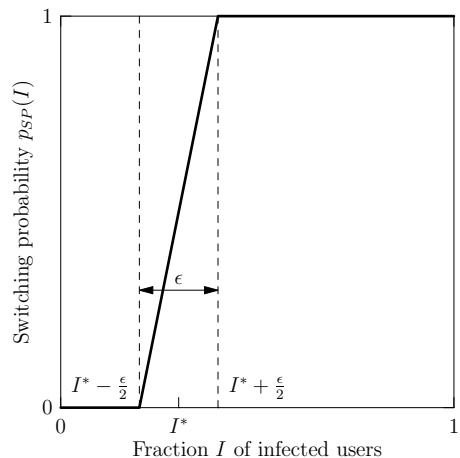}\\
  \caption{The user response function $p_{SP}(I)$ used in the simulations: The probability that a Susceptible user switches to being Protected, upon learning the fraction $I$ of Infected users in the network. }\label{fig:sigmoid}
\end{figure}

For the single user class case, we use a piecewise continuous response function (Figure~\ref{fig:sigmoid}):

\begin{equation}\label{eq:sigmoid}
p_{SP}(I) =
\begin{cases}
  0 & I < I^* - \frac{\epsilon}{2}\\
  \frac{1}{\epsilon}(I-I^*+\frac{\epsilon}{2}) & I^* - \frac{\epsilon}{2} < I < I^* + \frac{\epsilon}{2} \\
  1 & I > I^* + \frac{\epsilon}{2}
\end{cases}
\end{equation}
and $p_{PS}(I) = 1 - p_{SP}(I)$.

\begin{figure}[htbp]
  \centering\includegraphics[width=\textwidth]{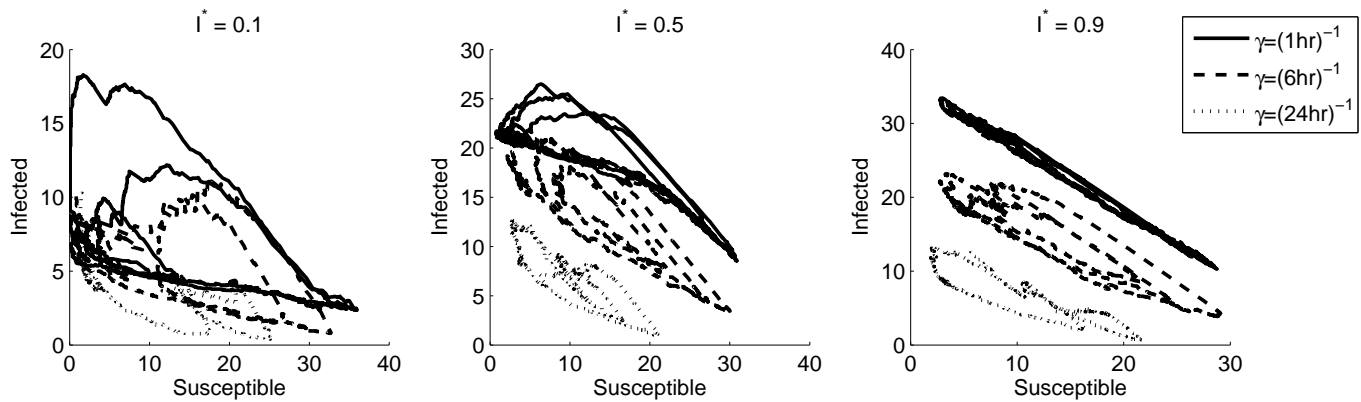}\\
  \caption{The trajectory of the system (average of 30 simulations) on the SI plane, when $\delta = (6hr)^{-1}$ and $\gamma$ takes the values $(1hr)^{-1}$, $(6hr)^{-1}$, and $(24hr)^{-1}$. The thresholds are $I^*=0.1, 0.5, 0.9$. We see that the network experiences higher numbers of Infected devices for higher values of $\gamma$, and for $I^*=0.1, 0.5$ we also observe the limiting effect of $I^*$.}\label{fig:single}
\end{figure}

In Figure~\ref{fig:single} we plot simulation results for $I^*=0.1, 0.5, 0.9$, and $\epsilon=0.001$, omitting an initial transient phase. Since $\beta(t)$ is not constant, the system state oscillates between two equilibrium points, $X_0$ (when $\beta(t)$ is low enough that $\delta>\beta(t)$) and either $X_1$ or $X_2$, depending on whether \eqref{eq:X1cond} is satisfied or not $\left(\frac{1-\frac{\delta}{\beta}}{1+\frac{\delta}{\gamma}} < I^*\right)$. Despite these periodicities, we see that for increasing values of $\gamma$ the system trajectories go through higher values of $I$, thus confirming our main conclusion that the infection level increases with the update rate. The effect of lowering $I^*$ is that it limits the maximum infection at the equilibrium, so the trajectories are capped at values of $I$ not far above $I^*$. For lower values of $I^*$, we see that the effect of $\gamma$ on the Infected is less pronounced.

For the two-user class case, we use separate piecewise continuous response functions for each class. Users of class 1 (Responsible) have a threshold of $I^{*1}=0.1$ and users of class 2 (Selfish) have a threshold of $I^{*2}=0.9$. For both classes $\epsilon=0.001$.

\begin{figure}[htbp]
  \centering
  \includegraphics[width=\textwidth]{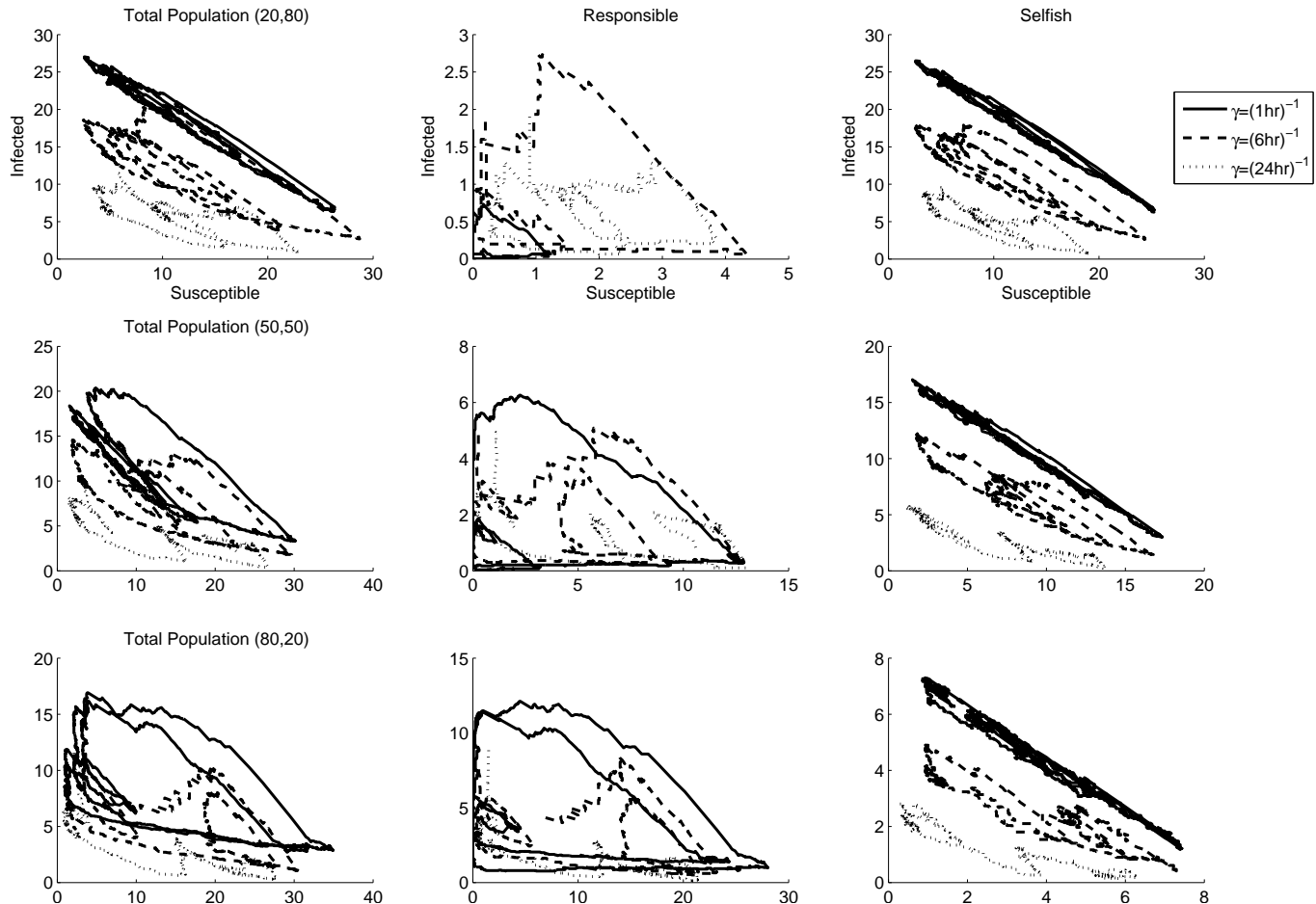}\\
  \caption{The trajectory of the system (average of 30 simulations) on the SI plane, when $\delta = (6hr)^{-1}$ and $\gamma$ takes the values $(1hr)^{-1}$, $(6hr)^{-1}$, and $(24hr)^{-1}$. Users are split into two classes: the Responsible, with $I^*=0.1$, and the Selfish, with $I^*=0.9$. The columns correspond to the Total Population, the Responsible subpopulation, and the Selfish subpopulation. The rows correspond to a total population split of 20\%-80\%, 50\%-50\%, and 80\%-20\% into Responsible and Selfish. We see, as in the case of a single user class, that the network experiences higher numbers of Infected devices for higher values of $\gamma$. In the current case of multiple user classes, the higher number of Infected is mostly due to the Selfish users.}\label{fig:multiple}
\end{figure}

In Figure~\ref{fig:multiple} we plot the system trajectories, again omitting an initial transient phase, for the Susceptible and Infected of 1) the total population (first column), 2) the Responsible subpopulation (second column), and 3) the Selfish subpopulation (third column). Each row corresponds to a different split of the total population into Responsible and Selfish subpopulations. In the first row, the Responsible-Selfish split is 20\%-80\%, in the second row it is 50\%-50\%, and in the third row it is 80\%-20\%.

We again confirm the conclusion that the fraction of Infected in the total population increases for larger values of $\gamma$. Two secondary conclusions relate to the situation within each subpopulation: The Selfish user trajectories seem as if the Selfish were isolated. That is, their trajectories are very similar to those they would follow if they were alone in the network (compare with the case $I^*=0.9$ in Figure~\ref{fig:single}). The Responsible users, on the contrary, stay mostly in the bottom left region, which means that many of them stay Protected. Comparing with the case $I^*=0.9$ in Figure~\ref{fig:single}, we see that they now stay a bit closer to the bottom left corner: This means that the Selfish-caused infection keeps more of them Protected than if they were alone in the network. The observations on the Selfish and on the Responsible are mutually compatible, as users that are Protected (here, the Responsible) do not interact with the rest of the network, so the trajectories of the remaining users (here, the Selfish) seem as if they were isolated.

\section{Conclusions}\label{sec:conclusions}
We have studied the effect of network users being cost-sensitive when deploying security measures. In particular, if users increasingly deploy security upon learning that the level of network infection is higher, and retract the deployment when the level of infection drops, then a higher learning rate leads to a higher equilibrium level of infected users.

We reach this same conclusion in three scenarios. Our main scenario is when users are strictly rational cost minimizers, having a discontinuous multi-valued best response behavior. The conclusion does not change when the response function is an arbitrary continuous single-valued function, as long as the function implies that users increasingly choose protection as the level of infection rises.  Finally, the conclusion remains valid even when there are two classes of users, each class with a different threshold function. These scenarios are studied both theoretically, by using a system of differential inclusions or differential equations, and also they are validated with simulations on human mobility traces.

We use the theory of differential inclusions to prove properties (existence, uniqueness, stability) of the system trajectories in the case of multivalued response functions. In the case of uniform user behavior, either continuous or discontinuous, the system is two-dimensional, and we are able to exclude the existence of periodic trajectories and to characterize the domains of attraction for each equilibrium point.

\end{document}